\documentclass[useAMS,usenatbib,preprint2]{aastex} 

\usepackage[landscape]{geometry}
\usepackage{graphicx}
\usepackage{times}
\usepackage{aas_macros} 
\usepackage{soul} 
\usepackage{bm} 
\usepackage{amssymb}
\usepackage{amsmath}
\usepackage{color}

\usepackage{booktabs} 

\input epsf

\newcommand\Rsq{{\mathbb{R}^2}}
\newcommand{\dif}{\mbox{$\mathrm{d}$}}
\newcommand{\um}{$\mu$m}

\newcommand{\Cfig}[2]{
    \begin{figure}
    \begin{center}
    \sfig{#1.eps}{0.9\columnwidth}
    \caption{{\small #2}}
    \label{fig:#1}
    \end{center}
    \end{figure}
}

\usepackage{suffix}

\newcommand{\vect}[1]{\bm #1}  

\newcommand{\dint}[2]{{\rm d}^{#1}#2\;}

\newcommand{\eeqp}{\;.\end{equation}}
\newcommand{\eeqc}{\;,\end{equation}}

\newcommand\refeq[1]{Eq.~(\ref{eq:#1})}
\WithSuffix\newcommand\refeq*[1]{(\ref{eq:#1})}

\newcommand\reftab[1]{Table \ref{tab:#1}}

\newcommand\reffig[1]{Figure \ref{fig:#1}}
\newcommand\reffigs[2]{Figures \ref{fig:#1}--\ref{fig:#2}}

\newcommand\refsec[1]{\S \ref{sec:#1}}
\newcommand\refapx[1]{Appendix \ref{sec:#1}}

\newcommand{\sfig}[2]{\includegraphics[width=#2]{#1}}

\slugcomment{\copyright\, Copyright 2013. All rights reserved.  To appear in PASP.}
\shorttitle{Weak Lensing Systematics from Image Combination}
\shortauthors{C. Shapiro et al.}

\begin{document}

\title{Weak Gravitational Lensing Systematics from Image Combination}

\author{
C.~Shapiro\altaffilmark{1}{$^\dagger$},
B.~T.~P.~Rowe\altaffilmark{2,1},
T.~Goodsall\altaffilmark{1},
C.~Hirata\altaffilmark{3,4},
J.~Fucik\altaffilmark{3},
J.~Rhodes\altaffilmark{1,3},
S.~Seshadri\altaffilmark{1},
R.~Smith\altaffilmark{3}
}
\email{$^\dagger$Charles.A.Shapiro@jpl.nasa.gov}
\email{ }

\altaffiltext{1}{Jet Propulsion Laboratory, California Institute of Technology, 4800
Oak Grove Drive, La Ca\~nada Flintridge, CA 91109}
\altaffiltext{2}{Department of Physics \& Astronomy, University College London,
Gower Street, London WC1E 6BT, UK}
\altaffiltext{3}{California Institute of Technology, 1200 East California
Boulevard, Pasadena, CA 91125}
\altaffiltext{4}{Department of Astronomy, Ohio State University, 4055 McPherson Laboratory, 140 West 18th Avenue, Columbus, OH 43210}

\date{ }
\date{Version as of \today. Copyright 2013. All rights reserved.  To appear in PASP.}


\label{firstpage}

\begin{abstract}

Extremely accurate shape measurements of galaxy images are needed to probe dark energy properties with weak gravitational lensing surveys.  To increase survey area with a fixed observing time and pixel count, images from surveys such as the Wide Field Infrared Survey Telescope (WFIRST) or Euclid will necessarily be undersampled and therefore distorted by aliasing.  Oversampled, unaliased images can be obtained by combining multiple, dithered exposures of the same source with a suitable reconstruction algorithm.  Any such reconstruction must minimally distort the reconstructed images for weak lensing analyses to be unbiased.  In this paper, we use the IMage COMbination (IMCOM) algorithm of Rowe, Hirata, and Rhodes to investigate the effect of image combination on shape measurements (size and ellipticity).  We simulate dithered images of sources with varying amounts of ellipticity and undersampling, reconstruct oversampled output images from them using IMCOM, and measure shape distortions in the output.  Our simulations show that IMCOM creates no significant distortions when the relative offsets between dithered images are precisely known.  Distortions increase with the uncertainty in those offsets but become problematic only with relatively poor astrometric precision.  E.g.\ for images similar to those from the Astrophysics Focused Telescope Asset (AFTA) implementation of WFIRST, combining eight undersampled images (sampling ratio $Q=1$) with highly pessimistic uncertainty in astrometric registration ($\sigma_d\sim10^{-3}$ pixels) yields an RMS shear error of $O(10^{-4})$.  Our analysis pipeline is adapted from that of the Precision Projector Laboratory -- a joint project between NASA Jet Propulsion Laboratory and Caltech which characterizes image sensors using laboratory emulations of astronomical data.

\end{abstract}


\section{Introduction}

\subsection{Background}\label{sec:bg}

Weak gravitational lensing is advancing rapidly as a tool for learning
about the dark Universe.  In particular, measurements of cosmic shear
-- the large-scale spatial correlation of galaxy shapes -- are
expected to yield tight constraints on the properties of dark matter
and dark energy \citep{Albrecht:2006uq}.  Several groups first
detected cosmic shear at the turn of the century using thousands of
galaxies distributed over small patches of sky
(\citealp*{baconetal00,kaiseretal00};
\citealp{vanwaerbekeetal00,wittmanetal00}).  Cosmic shear surveys grew
to include millions of galaxies (see, e.g., \citealp{hoekstrajain08}
for a recent review), with the largest survey to date being CFHTLenS
at about 6 million galaxies over 154 deg$^2$
\citep{heymansetal12,Kilbinger:2013qy}.  Surveys coming online now
will map hundreds of millions of galaxies over thousands of deg$^2$
(e.g.\ \citealp{sanchez10,miyazakietal12,dejongetal13}), while future surveys such as Euclid and
the Large Synoptic Survey Telescope (LSST) will push galaxy counts
into the billions \citep{laureijsetal11,abelletal09}.  Furthermore, the NASA's planned Wide Field
Infrared Survey Telescope (WFIRST) -- particularly in its 2.4m Astrophysics Focused Telescope Asset (AFTA) implementation -- offers a chance to create
exceptional weak lensing maps from its combination of high galaxy
density, a high median redshift and unparalleled systematic control \citep{gehrels10,dressleretal12,Spergel:2013ys}.

The shrinking statistical uncertainty of cosmic shear surveys is putting more stringent 
requirements on the error tolerances of the telescope and analysis methods.  A typical 
gravitational shear -- the anisotropic dilation and contraction of a galaxy's observed shape -- 
is O(10$^{-2}$) and adds to a galaxy's intrinsic ellipticity which is O(10$^{-1}$).  The correlation 
between galaxy pairs due to gravity alone is therefore O(10$^{-4}$), while the intrinsic shape 
noise is O(10$^{-2}$).  In order to avoid biasing measurements of dark energy properties and 
other cosmological parameters, systematic errors in the correlation measurements must be kept to 
O(10$^{-7}$) or smaller \citep{amararefregier08}.  Simple arguments show
that to achieve this requires knowledge of the size and ellipticity of the instrumental point-spread
function (PSF) to better than one part in $10^3$ \citep{paulinetal08,paulinetal09,masseyetal13}.
This unprecedented level of accuracy is causing optics, 
image sensors, and image processing algorithms to come under close
scrutiny.  

In this paper, we focus on errors that can arise from combining star and galaxy
images prior to shape measurement.  Image combination is often
necessary to overcome aliasing, which occurs when the pixel spacing on
a telescope's image sensor undersamples the full range of spatial
frequencies admitted by the optics.  Space missions in particular,
such as WFIRST-AFTA and Euclid, will produce undersampled, and therefore
aliased images at native resolution due to the lack of atmospheric
seeing, and a justifiable scientific desire to maximise field of view.
Aliasing can be overcome by taking dithered exposures that allow the
pixels to sample different parts of the source images, which include
the PSF.  The multiple exposures can then be used to reconstruct
oversampled (i.e.\ better than Nyquist-sampled) images.  In this work
we study the effect that this process has on shape measurement
(see, e.g.,
\citealp{lauer99,lauer99psf,fruchterhook02,rhodesetal07,fruchter11};
\citealp*{roweetal11}).  The most important consideration is the
propagation of defects due to aliasing into the higher resolution
image reconstructions \emph{even while dithering}: this can occur to
some degree for a range of image combination algorithms when applied
in the general case.  Where this can be overcome, however, a secondary
consideration is control over unwanted changes in the PSF.  Additional
filtering of the output image is a common result of interpolative
prescriptions for combining dithered images.

Multiple image combination algorithms have been suggested in the
literature (see
\citealp{lauer99,fruchterhook02,bertinetal02,fruchter11}; also
\refsec{aliasing}), but we restrict our attention to the IMCOM
algorithm of \citet{roweetal11}, which was developed for WFIRST-AFTA.  The advantages of IMCOM lie in its
generality, and in the qualitatively enhanced degree of control it
provides over the properties of oversampled output images, including
over aliasing.  Rather than applying a fixed interpolation scheme, it
solves for the optimal linear transformation between the pixels of the
input images and the pixels of the output image.  The optimization is
subject to user-specified noise or error tolerances, and it can
incorporate PSFs and grid distortions that vary from image to image.
IMCOM is described in further detail in \refsec{theory}.  Our goal is to
quantify shape distortions induced when IMCOM is applied to
undersampled data.  
We do not make any claims about the effectiveness of alternative algorithms.

IMCOM has become an integral
part of our Precision Projector Lab (PPL) experiments to assess the
impact of detector-induced shape distortions on weak lensing
measurements.  The PPL is a joint project by Caltech Optical
Observatories (COO) and NASA Jet Propulsion Laboratory (JPL).  Our
principal instrument -Ð an Offner-based re- imaging system
(a.k.a.\  Òthe projectorÓ) -Ð casts precisely controlled images onto
CCD, CMOS or IR detectors.  Measuring these images allows us to
characterize detectors and quantitatively understand their
non-idealities.  The projector can also emulate astronomical data such
as stars, galaxies, or spectra.  The original intent of the PPL, which
began in 2008, was to simulate galaxy shape measurement for a weak
gravitational lensing survey with the Joint Dark Energy Mission (JDEM).  
While continuing our weak lensing investigations for WFIRST-AFTA, we have also expanded our infrastructure and personnel to address other astronomical
applications such as spectrophotometery of transiting exoplanets with the James Webb Space Telescope \citep{Beichman:2012ly} and testing the NIR natural guide star sensor for the Keck-1 adaptive optics system \citep{Adkins:2012zr}.  
The projector and our initial measurements of detector-induced bias in the shear correlation function are described in more detail in
companion papers \citep[Smith et al.\ in prep]{Seshadri2013}.  These measurements rely on our findings described in this paper -- that any PSF distortions introduced by IMCOM are well
below those of our projector system and detector.

\subsection{Outline of Paper}
The goal of this paper is to demonstrate that reconstructing Nyquist-sampled 
images from aliased data with IMCOM introduces negligible distortions that have a minor impact on cosmic shear analyses.  
In \refsec{theory}, we review sampling theory, aliasing, and the theory behind IMCOM.  
In \refsec{analysis}, we describe our data analysis pipeline, including how IMCOM is applied.  
In \refsec{sim}, we describe our image simulations.  
In \refsec{tests}, we present the results of IMCOM's performance on several shape measurement tests.  
We conclude in \refsec{conclusions}.

\section{Review of Sampling Theory and IMCOM} \label{sec:theory}

\subsection{Nyquist sampling and aliasing in astronomical imaging}\label{sec:aliasing}
In the analysis of astronomical images, the adequate spatial
sampling of data by pixels of finite size and spacing is an important
consideration. 
Ideally, images used for science should be sampled at or above the
Nyquist-Shannon sampling rate for the \emph{band limit} set by the
optical response of the system (see e.g.\ \citealp{marks09}), so that
the full continuous image can be determined from the discrete pixel
samples.  If an image contains
only Fourier modes whose spatial frequency ${\bf u}$ is no larger in
magnitude than some bandlimiting frequency $u_{\rm max}$,
then the Nyquist criterion demands sample spacing $p$ satisfying
$p<1/(2u_{\rm max})$.  An image sampled more
finely than the critical rate $1/(2u_{\rm max})$ is referred to as
\emph{oversampled}; one sampled at the critical rate is
\emph{critically sampled}. It is convenient to define the sampling factor, 
\begin{equation}\label{eq:Qdef}
Q\equiv \frac{1}{p\,u_{\rm max}}
\end{equation}
so that the critical sampling condition becomes $Q=2$.
\emph{Undersampled} images for which $Q<2$ are
subject to aliasing, where spatial frequency modes spuriously appear as lower frequency distortions to
the original image. When $Q\le 1$, all modes in the image are aliased, which is referred to as \emph{strong undersampling}.

The spatial bandlimit of an astronomical image is
\begin{equation}
u_{\rm max} = \frac{1}{\lambda_{\rm min} N_f}
\end{equation}
where $N_f$ is the focal ratio or ``f-number'' of the telescope and $\lambda_{\rm min}$ is the shortest wavelength of the incident light.  Hence,
\begin{equation}
Q = \frac{\lambda_{\rm min}N_f}{p}  \;.
\end{equation}
For example, WFIRST-AFTA would be an $f/7.8$ instrument using IR detectors for imaging with pixel spacing $p=10$\um.  It has $\lambda_{\rm min}=1.38$\um\ in the H-band, thereby creating images with $Q=1.08$.  Euclid will be $f/20.4$ and use CCDs with $p=12$\um\ for imaging.  In its wide visible band, it has $\lambda_{\rm min}=0.55$\um, thereby creating images with $Q=0.94$.  Thus, every frequency in a Euclid image and nearly every frequency in a WFIRST-AFTA image will be aliased.

Because an aliased image has ``missed'' high frequencies due to an insufficiently small pixel spacing, the sampled image's geometric properties, such as centroid or ellipticity, will depend on its location relative to the pixel centers.  The precise shape of an aliased image cannot be recovered through interpolation or other image processing techniques.
\citet{fruchter11} gives an account of aliasing in astronomical images from the \emph{Hubble Space Telescope}, but it is problematic wherever images are undersampled.
By contrast, oversampled images allow full and accurate reconstruction of the underlying PSF-convolved image by sinc interpolation and derived quantities will be free
from aliasing defects.  This makes oversampled data the
preferred input for most precision image analysis applications,
including weak gravitational lensing where the morphological
distortions introduced by aliasing are potentially damaging \citep{Rhodes:2007fk}.  
Where oversampled images are not available at the native instrument resolution, one strategy is to attempt to combine multiple undersampled images to generate a synthetic oversampled image.

\subsection{Oversampled image generation with IMCOM}\label{sec:imcom}
As described in \citet{roweetal11} and \citet{cropperetal12}, IMCOM can be
used to reconstruct oversampled images from undersampled data with
precise control over both noise in the final image and unwanted
distortions to the PSF.  Standard algorithms for
combining multiple images, which use interpolation-like recipes for allocating the
flux in the input pixels to the output image (e.g.\ DRIZZLE: \citealp{fruchterhook02};
SWARP: \citealp{bertinetal02}),
 typically introduce additional distortions to the image (in all but a subset of symmetric special cases). 
These distortions, which may be described as an unwanted change to the PSF of the image, will depend on the precise recipe used for image combination and the configuration of input and output pixels.  
They are therefore not easily controlled in general.

The IMCOM algorithm avoids these issues by making control over the PSF one of two
metrics used to determine the optimal linear combination of input
pixels; the second metric is noise in the output image.  In this way
IMCOM differs qualitatively from other methods, and wherever other
methods are linear they fall within the search space of possible combinations.  We now
briefly describe the IMCOM method (for a more detailed description see
\citealp{roweetal11}).

The undersampled input images are written as a vector of intensities 
$I_i$ of length $n$, where $n$ is the total number of usable
pixels. We describe the PSF at each pixel location $\textbf{r}_i$, including all convolutive
effects such as image motion, optics, and detector response, as the function $G_i({\bf r})$.  The intensity at the
$i$th pixel is thus given by  
\begin{equation}
I_i = \int_\Rsq f({\bf r}') G_i({\bf r}_i-{\bf r}') \dif^2{\bf r}' + \eta_i,
\label{eq:I}
\end{equation}
where the function $f({\bf r}')$ describes the physical image on the
sky and $\eta_i$ is the noise with $\langle\eta_i\rangle=0$ and some
covariance matrix $N_{ij}=\langle\eta_i\eta_j\rangle$. The formalism
allows any general noise covariance matrix $N_{ij}$, although in most
cases $N_{ij}$ is close to diagonal (off diagonal terms may be
introduced by, e.g., inter-pixel coupling, $1/f$ noise).  

Let us then seek an output image $H_\alpha$ on a grid of pixel centers
${\bf R}_\alpha$, where $\alpha=1, \dotsc, m$.  Typically we will
choose ${\bf R}_\alpha$ so the output image $H_{\alpha}$ is
oversampled.  Note that we will follow the notation of
\citet{roweetal11} in which Latin indices are used for input pixel
locations, and Greek indices are similarly used to refer to output
pixel locations.  The general linear expression for $H_\alpha$ in
terms of $I_i$ is simply
\begin{equation}\label{eq:ha}
H_\alpha = \sum_i T_{\alpha i} I_i,
\end{equation}
where $T_{\alpha i}$ is an $m\times n$ matrix.  The IMCOM algorithm
is one prescription for finding the $T_{\alpha i}$ that gives an $H_\alpha$
with optimal properties under a chosen objective function.  

The objective function chosen by \citet{roweetal11} relates $H_\alpha$ to
a desired \emph{target image} $J_{\alpha}$, which is defined as:
\begin{equation}\label{eq:Ja}
J_\alpha \equiv \int_\Rsq f({\bf r}') \Gamma({\bf R}_\alpha-{\bf r}') \dif^2{\bf r}'.
\end{equation}
Here $\Gamma$ is the desired effective PSF of the synthesized output
image.  A well-motivated choice for $\Gamma$ is often simply the input 
$G_i({\bf r})$ (if this is approximately constant between input pixels) so that $H_{\alpha}$ 
contains no 
unwanted additional contributions to the PSF.  This will be discussed
further in \refapx{ggamma}.

How the relationship between the target image $J_\alpha$ and $H_\alpha$
is used to construct a suitable objective function is described in detail
in \citet{roweetal11}, but we now present a summary of the
prescription developed in that paper.
The objective function is made
from two terms, linked by a Lagrange multiplier $\kappa > 0$. 
The first term is the \emph{leakage objective}
$U_{\alpha}$, given by
\begin{equation}\label{eq:Ulong}
U_{\alpha} = \int_\Rsq  \left[L_{\alpha} ({\bf r})\right]^2 \dif^2 {\bf r},
\end{equation}
for a leakage function defined as
\begin{equation}\label{eq:L}
L_{\alpha} ({\bf R}_{\alpha} - {\bf r'}) \equiv  \sum_i T_{\alpha i} G_i({\bf r}_i-{\bf r}') - 
\Gamma({\bf R}_\alpha-{\bf r}').
\end{equation}
The leakage function is simply the difference between the desired PSF
$\Gamma$ and its actual reconstructed counterpart in $H_{\alpha}$, and
by minimizing $U_{\alpha}$ this difference is minimized in a
least-squares sense. (Note that in the expression for $U_{\alpha}$
above we have explicitly chosen $\Upsilon({\bf r})$ to be a Dirac
delta function; see \citealp{roweetal11}.)

The second term is $\Sigma_{\alpha \alpha}$, the diagonals of the
noise covariance matrix for the output image as given by
\begin{equation}
\Sigma_{\alpha\beta} = \sum_{ij} T_{\alpha i}T_{\beta j}N_{ij}.
\label{eq:Cr}
\end{equation}
As in \citet{roweetal11}, we do not adopt the Einstein summation
convention for repeated indices.  Together with the Lagrange
multiplier $\kappa_{\alpha}$, the overall objective function adopted in IMCOM
is then
\begin{equation}
W_\alpha = U_{\alpha} + \kappa_{\alpha} \Sigma_{\alpha\alpha}.
\label{eq:Wa}
\end{equation}
Because equation \eqref{eq:Wa} involves quadratic combinations of $T_{\alpha i}$, its
derivative with respect to $T_{\alpha i}$ is linear, and minimizing $W_\alpha$ uniquely
 determines a solution $T_{\alpha i}(\kappa_\alpha)$.  For a user-specified tolerance on the 
 output noise variance ($\Sigma_{\alpha \alpha}$) or on unwanted distortion in the
 output image (characterized by $U_{\alpha}$), IMCOM solves for the values of $\kappa_{\alpha}$ 
 which provide a $T_{\alpha i}$ that minimize $W_\alpha$ while satisfying the tolerance.

\subsection{Tolerances on $U_{\alpha}$ }\label{sec:u7u8}
As for the tests performed by \citet{roweetal11} and \citet{cropperetal12}, we will use the IMCOM software to construct solutions $T_{\alpha i}$ for the image combination that satisfy a threshold on the maximum tolerated value of the leakage objective $U_\alpha$. We label this maximum $U^{\rm max}_{\alpha}$.  IMCOM then seeks a solution for the values of the Lagrange multiplier $\kappa_{\alpha}$ which provide a $T_{\alpha i}$ that satisies $U_{\alpha} < U^{\rm max}_{\alpha}$ at the location of each output pixel $\textbf{R}_{\alpha}$ and minimizes $\Sigma_{\alpha \alpha}$ \citep{roweetal11}.

In both these previous studies, the threshold adopted for $U_{\alpha}$ was $U^{\rm max}_{\alpha} = 10^{-8} C_{\alpha}$, where
\begin{equation}\label{eq:Ca}
C_{\alpha} = \int_\Rsq  \left[\Gamma_{\alpha} ({\bf r})\right]^2 \dif^2 {\bf r}
\end{equation}
which, as for equation \eqref{eq:Ulong} relating $U_{\alpha}$ to the leakage $L_{\alpha}$, is contingent on the choice of $\Upsilon({\bf r})$ as a Dirac delta function (see \citealp{roweetal11}).  From equations \eqref{eq:Ulong} \& \eqref{eq:Ca} the value of $U^{\rm max}_{\alpha} / C_{\alpha}$ can be seen intuitively as placing a requirement on the integrated, squared leakage $L_{\alpha}({\bf r})$ relative to the integrated, squared target PSF, $\Gamma_{\alpha}({\bf r})$.  In a root mean square (RMS) sense, choosing $U^{\rm max}_{\alpha} / C_{\alpha} = 10^{-8}$ could be described as limiting unwanted changes in the PSF to be smaller than one part in $10^4$.

Given the O($10^{-3}$) requirements for the knowledge of the size and ellipticity of the PSF for a successful weak lensing experiment (see \refsec{bg}; \citealp{paulinetal08,paulinetal09,masseyetal13}), $U^{\rm max}_{\alpha} / C_{\alpha} = 10^{-8}$ represents a conservative choice. This low tolerance value ensures that any contributions to the PSF uncertainty budget will be small in all but the most contrivedly pathological (and unrealistic) cases.  However, $U^{\rm max}_{\alpha} / C_{\alpha} = 10^{-8}$ is a stringent condition requiring excellent sampling of the image plane from the input dither configuration $\textbf{r}_i$.  As each additional dithered exposure that contributes to $\textbf{r}_i$ is costly in terms of survey depth, total exposure time, or both, it is appropriate to ask whether the value of $U^{\rm max}_{\alpha} / C_{\alpha}$ may be relaxed somewhat, while ensuring that unwanted distortion in the PSF be kept tolerably small.

We undertake a first investigation of this question by relaxing our tolerance on $U_{\alpha}$ to $U^{\rm max}_{\alpha} / C_{\alpha} = 10^{-7}$ in the analysis presented in this paper.  This limits RMS unwanted distortions to the PSF to $\sim$3 parts in $10^4$.  Intuitively, and in the absence of avoidable asymmetries in the offset patterns of dithered exposures, RMS distortions of this order would not normally be expected to produce comparable or larger changes in the PSF size and ellipticity (the PSF characteristics of leading order importance in weak lensing).  This suggests that cosmic shear requirements ought still to be met comfortably even while relaxing to $U^{\rm max}_{\alpha} / C_{\alpha} = 10^{-7}$.  The results of our tests in \refsec{tests} demonstrate that this is indeed the case.

\section{Analysis Pipeline} \label{sec:analysis}

\subsection{Summary of PPL pipeline} \label{sec:pipeline}

We test IMCOM by running a portion of the PPL analysis pipeline on simulated images.  The pipeline was designed to reduce controlled 
image data for our investigation of detector-induced shape measurement errors.  Here, we 
summarize that pipeline in order to put IMCOM usage in context.  The pipeline contains essential 
steps which would be part of a more sophisticated analysis of real science images.

The following pipeline steps are typically required in our non-simulated data:
\newcounter{saveenum}
\begin{enumerate}
\item Sources in each calibrated, undersampled image are detected with SExtractor \citep{Bertin:2010fk}, which 
provides centroid estimates.  
\item Sources are tracked across multiple exposures.  If a source cannot be found in one image 
(not detected, moves off edge) or is within 5 pixels of a known bad pixel, the object is removed from all catalogs throughout the rest 
of the analysis.
\item During data acquisition, sequences of dithered exposures are taken by translating the source pattern transversely with respect to the optical axis.  
The relative $(x,y)$ positions of the exposures are measured by computing the change in the average centroid position of all the sources.  
Noise in the centroid estimates is attenuated by averaging over many sources (typically several thousand).  
The dithers consist of random translations for reasons discussed in \refsec{multistar}
\setcounter{saveenum}{\value{enumi}}
\end{enumerate}
In this paper, we bypass the above steps by simulating both the dithered images and the measured dither positions (relative coordinates of the images).  By adding varying amounts of random error to the (precisely known) true dither positions, we can estimate how shape distortions in the reconstructed images depend on astrometric error.  Shapes are derived from the following steps:
\begin{enumerate}
  \setcounter{enumi}{\value{saveenum}}
\item The dither positions are supplied to IMCOM, along with the 
required PSF models and user-defined ``soft'' parameters \citep[as defined in][]{roweetal11}.  IMCOM finds the optimal solution for 
reconstructing output images. 
\item Small sub-images or ``postage stamps'' of the sources are extracted from the input 
exposures.  IMCOM applies its solution to create an oversampled sub-image image of each source.
\item The shapes of the oversampled sources are measured by computing combinations of their second moments (see \refsec{moment}).
\end{enumerate}
In \refsec{tests}, we will show that the largest source of shape measurement error caused by this pipeline comes not from IMCOM itself but from the astrometric errors.

\subsection{Ellipticity Moment Measurement on Oversampled Images} \label{sec:moment}

Ellipticity is commonly used as an estimator for shear in weak gravitational lensing analyses. Note that these are not synonymous concepts: ellipticity is a geometric property describing the light profile, while shear refers to a linear transformation applied to the galaxy image. Multiple conventions for calculating ellipticity can be found in the literature \citep{Bernstein:2002uq,Schneider:2006kx}. For the sake of conceptual and computational simplicity, we adopt the convention of computing ellipticity using weighted quadrupoles \citep{Kaiser:1995fk}.
If $I(\vect{r})$ is the intensity profile of the object, then weighted ellipticities, $e_1$ and $e_2$, are defined as:
\begin{equation}
e_1 \equiv \frac{M_{xx}-M_{yy}}{M_{xx}+M_{yy}} \hspace{1cm}
e_2 \equiv \frac{2M_{xy}}{M_{xx}+M_{yy}}
\end{equation}
\begin{equation}
M_{ij} \equiv \frac{ \int \dint{2}{r} I(\vect{r})w(\vect{r})(r_i-\bar r_i)(r_j-\bar r_j) }
					{ \int \dint{2}{r} I(\vect{r})w(\vect{r})}
\end{equation}
where $i$ and $j$ correspond to either axis of the pixelated image, and $\bar{\vect{r}}$ is the weighted image centroid (1st moment).  The weighting function $w(\vect{r})$ is introduced to ensure that the integrals converge in the presence of noise.  We take $w(\vect{r})$ to be a radially symmetric 2D Gaussian centered at $\bar{\vect{r}}$ (the width is discussed in \refsec{setup}).  The location of $\bar{\vect{r}}$ is found iteratively by calculating the weighted centroid then re-centering $w(\vect{r})$ on that centroid until the result changes by less than $10^{-4}$ pixels.  We compute the integrals by simply summing over pixel positions in the output postage stamp without interpolating, finding that any numerical error so introduced is negligible. 
Typically, one would estimate and subtract background noise before measuring shapes; however, we omit this step since we are not simulating noise.\footnote{A weighting function is not needed for a noise-free image either, but we include it so that our results can be compared to typical analyses.} 


Since error tolerances for weak lensing surveys are quoted in terms of gravitational shear, not ellipticity, we need a calibration to convert errors in the ellipticity estimator to inferred errors in the shear.  A realistic calibration would be a function of the source properties such as shape and signal-to-noise -- this is called ``shear susceptibility'' \citep[see e.g.~][]{Leauthaud:2007fk}.  In addition, errors in ellipticity estimates from stellar point sources must be carefully propagated to the shear errors in nearby galaxies \citep{paulinetal08}.  Such a full treatment of PSF deconvolution is beyond the scope of this paper.  For the purpose of estimating the effects of image combination on shape measurement, we find that a crude scaling factor is sufficient to relate our simulated measurements to weak lensing shear requirements.  We define our calibration factor as
\begin{equation}
P \equiv \frac{2\gamma_i}{e_i} \;.  \label{eq:calibrate}
\end{equation}
For each set of simulations we run, we determine the calibration factor experimentally by measuring ellipticities for oversampled images of sources with known shears.  The ellipticity measurement includes the simulated effects of the detector (charge diffusion, pixelization), and the Gaussian weighting function, both of which will have a rounding effect on the image.  The calibration factor therefore accounts for these effects to provide an estimate of the shear.  For our ellipticity definition, if we set $w(\vect{r})=1$, then a noise-free, radially symmetric image sheared by $\gamma_i$ will have $e_i=2\gamma_i$ and $P=1$ \citep{Kaiser:1995fk}.  
\reftab{simparams} lists our estimated $P$ for various simulation settings.  Note that as $Q$ decreases, the sizes of the input pixels are growing, causing an increasingly large blurring effect on the PSF that must be compensated by larger values of $P$.

We are also interested in measuring the weighted image size, defined as
\begin{equation}
R^2 = M_{xx}+M_{yy} \;.
\end{equation}
Distortions to image size are important in weak lensing analyses since the PSF must be accurately measured and deconvolved from galaxy images.  \citet{paulinetal08} showed that size-measurement errors in the PSF (from point sources) propagate to shear measurement errors in the galaxy according to
\begin{equation}
\Delta e_i \approx (e_i^{\rm galaxy} - e_i^{\rm PSF})\frac{\Delta R^2_{\rm PSF}}{R^2_{\rm galaxy}} \;.
\end{equation}
Measurements of $\Delta R^2_{\rm PSF}/R^2_{\rm PSF}$ therefore provide a rough estimate of the resulting multiplicative shear error for a typical weak lensing galaxy with $|e^{\rm galaxy}| \gg |e^{\rm PSF}|$ and $R^2_{\rm galaxy}\sim R^2_{\rm PSF}$.


\section{Simulations} \label{sec:sim}

\subsection{Simulating individual sources}\label{sec:sources}
We base our image simulations on a simple model PSF for a telescope, including the detector. 
For light at wavelength $\lambda$ incident upon a simple
circular pupil, for an optical system at focal ratio $f/N_f$, the diffraction pattern is described by the well known Airy spot,
\begin{equation}\label{eq:airy}
  I_{\rm optical}(x ; \lambda N_f) = \left[\frac{2 J_1 \left(\pi x / \lambda N_f \right)}{\left(\pi 
  x / \lambda N_f \right)} \right]^2 ,
\end{equation}
where $J_1$ is the first order Bessel function of the first kind, and
$x$ is the radial distance in the focal plane from the center of the
spot in the plane normal to the optical axis.  A real telescope will have a more complex optical PSF due to e.g. obscurations and aberrations.  For the purposes of testing image reconstruction, the most important feature of the optical PSF is the bandlimit (see \refsec{aliasing}).

The two dimensional Fourier transform of the PSF is the \emph{optical transfer function}, whose absolute value is the \emph{modulation transfer function} (MTF).  For a circularly symmetric
function such as $I(r)$, the MTF is also circularly symmetric, and
can be expressed via the zeroth order Hankel transform
\begin{equation}
\tilde{I}(u) = 2 \pi \int_0^{\infty} \! I(x)  J_0(2 \pi u x) {\rm d}x ,
\end{equation}
where $J_0$ is the zeroth order Bessel function of the first kind.
The MTF for the Airy disc $I_{\rm
  optical}(x; \lambda N_f)$ is then
\begin{equation}
\tilde{I}_{\rm optical} (u; u_{\rm max}) = \frac{2}{\pi} \left[ \arccos{\left(\frac{u}{u_{\rm
          max}} \right)}  -
  \frac{u}{u_{\rm max}} \sqrt{1 - \left(\frac{u}{u_{\rm max}}
    \right)^2 }\right],
\end{equation}
where $u_{\rm max} = 1 / \lambda N_f$.  This result can be derived by
considering that the MTF of the Airy disc is the autocorrelation of a
top hat function of radius $u_{\rm max} / 2$, and using a geometric
argument. 

We approximate the effects of charge diffusion in the
detector material by convolving $I_{\rm optical}$ with a circularly
symmetric Gaussian kernel $I_{\rm cd}$ given by
\begin{equation}
I_{\rm cd} (x; \sigma_{\rm cd}) = \frac{1}{2 \pi \sigma^2_{\rm cd}} \exp{\left(
      -\frac{x^2}{2 \sigma^2_{\rm cd}}\right)},
\end{equation}
where $\sigma_{\rm cd}$ is the rms distance covered in the detector focal plane
by randomly diffusing electrons.  The MTF for this charge diffusion profile is
given by
\begin{equation}
\tilde{I}_{\rm cd} (u; \sigma_{\rm cd}) = \exp{\left(- 2 \pi^2 \sigma^2_{\rm cd} u^2 \right)}.
\end{equation}

Detectors integrate the total number of electrons within pixel
boundaries, and provide the integrated value as the output flux.  It
can be simply shown that integrating image flux within ideal square
pixels is equivalent to: i) convolving with a square boxcar filter
\begin{equation}
I_{\rm pixel}(x_1, x_2; p/2)  = \left\{  \begin{array}{cc} 1 / p^2 & \max
    \left\{ |x_1|, |x_2| \right\} \le p \\
  0 & {\rm otherwise} \end{array} \right.
\end{equation}
where $p$ is the pixel width and $x_1$ and $x_2$ are perpendicular
distances in the plane normal to the optical axis in the focal plane
(so that $x^2 = x_1^2 + x_2^2$); and ii) sampling the resulting
convolved function at the locations of the pixel centres.  The MTF of
the boxcar filter function is given by
\begin{equation}
\tilde{I}_{\rm pixel} (u_1, u_2; p) = {\rm sinc}{ \left(u_1 p \right) } \: {\rm sinc}{
    \left(u_2 p \right)}, \label{eq:ftpixel}
\end{equation}
where  
${\rm sinc}{(x)} \equiv \sin{(\pi x)} / (\pi x) $, and where $u_1$ and $u_2$
correspdond to $x_1$ and $x_2$ in the Fourier domain, so that $u^2 =
u_1^2 + u_2^2$.

We can now construct an approximate model of the PSF by
convolving $I_{\rm optical}$, $I_{\rm cd}$, and $I_{\rm pixel}$ to
give a combined image, $I_{\rm PSF}$.
The convolution is simply expressed in the Fourier domain as:
\begin{eqnarray}
\tilde{I}_{\rm PSF}(u_1, u_2; u_{\rm max}, \sigma_{\rm cd}, p) & = &  
\tilde{I}_{\rm optical}\left( \sqrt{u_1^2 + u_2^2}; u_{\rm max}
\right)  \nonumber \\
& \times & \tilde{I}_{\rm
  cd}\left( \sqrt{u_1^2 + u_2^2}; \sigma_{\rm cd} \right)  \nonumber
\\
& \times &\tilde{I}_{\rm
  pixel} \left(u_1, u_2; p \right).\label{eq:ftsource}
\end{eqnarray}
Images of extended sources are simulated by convolving this PSF with a source model.  An exponential model of a galaxy profile is sufficient for our purposes.  The profile is
\begin{equation}
I_{\rm exp}(x; r_0) \equiv \frac{1}{2\pi r_0^2}  \exp\left(\frac{x}{r_0}\right) \;,
\end{equation}
and the Fourier Transform is
\begin{equation}
\tilde{I}_{\rm exp}(u; r_0) = (1+4\pi^2r_0^2u^2)^{-3/2} \,.
\end{equation}
The galaxy image model is then given by $\tilde{I}_{\rm source}=\tilde{I}_{\rm exp}\tilde{I}_{\rm PSF}$ in the Fourier domain.  Point source images simply have $\tilde{I}_{\rm source}=\tilde{I}_{\rm PSF}$ by definition.

Rendering a pixellized image of
the source is achieved by sampling $I_{\rm source}(x_1, x_2)$ at the pixel
centres (integration across pixels already having been accounted for
by $I_{\rm pixel}$).  Constructing a sample of $I_{\rm source}(x_1, x_2)$ on a uniform
pixel grid of spacing $p$ is straightforwardly achieved by first creating a uniform grid of
$\tilde{I}_{\rm source}(u_1, u_2)$ at
a spacing $\Delta u$ in Fourier space, and then using the
Discrete Fourier Transform.  Note that the real space image of the PSF formally extends to infinity and would require infinite frequency resolution to render perfectly.  The Fourier Transform of a finite resolution $\tilde{I}_{\rm source}$ is a regular grid of source images with period $1/\Delta u$ in real space, where neighboring images overlap.  The accuracy of ${I}_{\rm source}$ is therefore limited by $\Delta u$.


Elliptical extended sources are obtained by dilating and contracting $I_{\rm exp}$ along orthogonal axes with a simple coordinate transformation.  The scaling factor is
\begin{equation} \label{eq:scalingfactor}
s \equiv \left(\frac{2}{1-|e|} -1 \right)^{1/4}
\end{equation}
where $|e|$ is the magnitude of the ellipticity.  For example, an exponential profile with $e_1=|e|$ and $e_2=0$ is dilated by $s$ along the $x$-axis of the image and contracted by $1/s$ along the $y$-axis (note $s\ge 1$).  This affine transformation is straightforwardly related to the Fourier domain.  Once the elliptical profile is convolved with $I_{\rm PSF}$, the $|e|$ measured from the image will generally be smaller than the ``intrinsic'' ellipticity of the exponential profile.  We can also create elliptical PSFs by performing the same transformation on $I_{\rm optical}$ before rendering a point source image.  Here, we use \refeq{scalingfactor} to set $|e|$ for the Airy profile before convolving it with $I_{\rm cd}$ and $I_{\rm pixel}$. This has the side-effect of increasing the bandlimit along the contracted axis by a factor of $s$. 
The utility of elliptical PSF images is described in \refsec{setup} along with our simulation settings.

\subsection{Simulating Images for Analysis} \label{sec:multistar}

In order to quantify PSF distortions from IMCOM and related sources of shape measurement error, 
we simulate images which are easily processed by the current PPL analysis pipeline.  For 
our purposes, it is sufficient to render sources in simple grid patterns.  
As described below, we are able to significantly reduce computational overhead by applying the same IMCOM reconstruction solution to all the sources in a single grid.
Along one grid axis, we vary the ellipticity of the sources, and along the other axis, we simulate varying amounts of astrometric error.  
Our results do not depend on the size or layout of the grids.

To render a grid of sources, we create a blank image and tile it with small 
``postage stamp'' images, each containing a single source.  The postage stamps 
do not overlap, therefore no de-blending is needed.
The sizes of the postage stamps must be  sufficiently large to reduce edge-effects in the output images reconstructed by IMCOM.
Each output pixel value is given by a linear combination of \emph{all} input pixels, with significant contributions coming from input pixels that are physically within a few PSF widths of the output pixel.  Therefore, the input postage stamps should be larger than the desired output postage stamp by a border of a few PSF widths, typically 2 to 3.
The output postage stamp from IMCOM must also be large enough so that the numerical error in the 
calculation of second moments is sufficiently small.  
Using the simulation settings summarized in \reftab{simparams}, we find that we are computationally limited by the accuracy in the images themselves, set by the resolution by which the functions are generated in Fourier space.

The coordinates of source centroids within the postage stamps are randomized in order to average 
over pixelization effects, which are important when the simulated source is comparable to the size 
of a pixel.  For each source in one grid image, the centroid is given a uniform random 
translation over 1 square pixel from the center of its postage stamp.  With the relative 
positions of the centroids fixed, we then simulate dithered images of the same source grid by 
adding a common uniform random translation to all the centroids.  When we render $N_s$ sets of 
$N_d$ dithers of the grid, each set receives new random dithers and new random 
relative source positions.  We can save the simulated positions in a file which we feed to the pipeline, which results in IMCOM knowing the precise relative translations between each dithered image.  Since a set of dithers is common to the entire grid, IMCOM can apply the same reconstruction solution to each of the $N$ sources in the grid, which speeds up computation significantly.  To simulate the effects of astrometric errors, source images are given additional random translations so that their true centroids do not match those in the positions file.

The reason for simulating random dithers instead of a fixed pattern is that precise dithers cannot be obtained in practice.  For instance, the WFIRST-AFTA survey strategy will image each source using multiple detectors in the focal plane to overcome the loss of area due to gaps between individual imagers.  The scan pattern, combined with the varying plate distortion across detectors, will yield effectively random relative dither locations (modulo one pixel) for a given source.   When arranged in a fixed, ideal pattern, the number of dithers needed to reconstruct an oversampled image is $N_d=(\rm{floor}(2/Q))^2$.  For instance, a 2x2 grid of dithers spaced precisely 0.5 pixels apart can be used to reconstruct Nyquist-sampled or oversampled images when $Q=1$.  With randomized relative positions, more dithers are needed to ensure they contain enough information to achieve Nyquist-sampling.

The number of dithers needed for reconstruction is determined experimentally.  Running IMCOM produces a file that indicates whether the user-specified error tolerance (``leakage objective''; see \refsec{u7u8}) on the reconstructed PSF has been met.  Note that IMCOM solves for the optimal mapping from input pixels to output pixels, and its solution depends only on the pixel locations and PSF input - not on the pixel values of the images themselves.  Thus we can test dither patterns by running IMCOM on sets of $N_d$ dummy images along with randomly generated dither positions files.  We increase $N_d$ until we find that 200 consecutive, independent sets of random dithers have met the error tolerance.  This method works well in practice but doesn't guarantee that all future random dither sets will meet the tolerance; therefore, we monitor the IMCOM output in all tests to verify that reconstructions are successful.

\section{Distortion Tests}  \label{sec:tests}

\subsection{Simulation and Pipeline Setup} \label{sec:setup}

Our goal is to determine what PSF distortions are introduced by the image combination process in converting undersampled images into oversampled ones.  
Therefore we compare shape measurements on simulated, oversampled images which were generated with and without the use of IMCOM.  
For a given set of simulation settings, we create a reference model $I_{\rm source}$ as in \refsec{sources}, which accounts for diffraction, charge diffusion, and pixel response.  
This model is sampled at a pixel spacing $p_{\rm out}$ such that $Q\ge 2$, i.e. the image is critically sampled or oversampled.  
To create input images for IMCOM, the model is also sampled at the pixel spacing $p_{\rm in}$ of 
a fiducial detector such that $Q<2$.  Input images are rendered repeatedly with random offsets 
applied to the source centroids to simulate dithering (see \refsec{multistar}).  The dithered input images are run through the PPL pipeline, which uses IMCOM to convert them into an oversampled output image at the pixel spacing $p_{\rm out}$.  Note that reference image has the same 
sampling as the output image, $p_{\rm out}$, but has a pixel response function that matches the 
input images since this is set by the size of the detector pixels, $p_{\rm in}$.  We then measure 
differences in the the calibrated shear $(\gamma_1,\gamma_2)$ and the size $R^2$ between the output image and 
the reference image, where $\gamma_i$ is given by \refeq{calibrate}.  This procedure (``one simulation'') is repeated many times to average over the random dither 
positions.
We do not add simulated shot noise or other systematic effects to the input images since we want to quantify the distortions from image combination alone.

We expect shape measurement errors due to image combination to decrease as the size of an image grows relative to the size of the PSF.  This is because in the IMCOM mapping of input pixels to output pixels, contributions of the input pixels to the output are most significant within a PSF width.  Therefore, measuring distortions for reconstructed point sources establishes a worst-case scenario.  In practice, a weak lensing survey would contain many galaxies with light profiles comparable in size to the PSF (i.e.\ barely resolved), making the images slightly larger than the PSF itself\footnote{Forecasts for WFIRST-AFTA, Euclid and LSST use galaxies as small as 0.8 PSF widths \citep{Spergel:2013ys}.} \citep{Jouvel:2009qf,Miller:2013vn}.
Hence, it is constructive (and computationally convenient) to test IMCOM on simulated point source images by varying parameters in the PSF model in \refsec{sources}.  In $\refsec{extended}$, we repeat our analysis for extended sources to verify that image combination in that case results in smaller shape distortions.

The parameters for our simulations are intended to test IMCOM over a range of 
sampling factors $Q$.  
Note that there is quite a bit of freedom in the parameter choice since the image combination process depends primarily on $Q$.
We look to the specifications for WFIRST-AFTA as a guide for realistic parameters, however we do not specifically simulate WFIRST-AFTA or any other mission.  We take our telescope to have a focal ratio $f/10$ 
($N_f = 10$ in Eq.\ \ref{eq:airy}), and we assume image sensors with a pixel pitch of $p_{\rm in}=$10\um. 
With these parameters, we simulate monochromatic images at various source wavelengths $\lambda_*$ to achieve the desired $Q$.  The images, which contain sources with ellipticities up to $|e_{\rm max}|=0.3$, have $Q=N_f\lambda_*/(p_{\rm in}s_{\rm max})$, where $s_{\rm max}=1.17$ accounts for the increased bandlimit due to our method for generating elliptical PSF images (see \refsec{sources}).
To approximate the effects of charge diffusion in the image sensor, we convolve images with a Gaussian filter of width $\sigma_{\rm cd}=2.9$\um, 
corresponding to a detector with charge diffusion length $\ell=1.87$\um\;  \citep[as measured for Teledyne Hawaii-2RG infrared detectors; see][]{Barron:2007bh,Seshadri2013}.  
Since the full-width-half-maximum (FWHM) of an Airy spot is given by FWHM $\approx 1.03\lambda_*N_f$, the point source images are roughly $Q$ input pixels wide (FWHM $\sim Qp_{\rm in}$) before shearing, pixelization, and charge diffusion are included.  Oversampled images of the point sources are roughly 2 pixels wide (FWHM $\sim 2p_{\rm out}$) since the output pixel sizes are defined such that $Q=2$ or marginally greater.  The output image shapes are always calculated including a Gaussian weighting function with $\sigma=4.5p_{\rm out}$, or $\sim$2.25 times the PSF width.

In addition to the input images and their relative coordinates, IMCOM requires models for the system PSF $G_i$ and target PSF $\Gamma$ (see \refsec{imcom}).  The user has some freedom in specifying these, but the reconstructed images will only be truly optimal if $G_i$ closely matches the PSF of the input images.  Choosing an appropriate PSF is described further in \refapx{ggamma}.
The most important aspect of the PSF models is that they have a spatial bandlimit greater than or equal to those of the images, otherwise spatial modes will be incorrectly removed from the output images.  Since the bandlimits of our simulated elliptical images are increased by a factor of $s$ relative to ordinary point sources, we compensate by using PSF models with $\lambda_{\rm PSF}=\lambda_*/s_{\rm max}$.
A set of images with a given $\lambda_*$ therefore has $Q=N_f\lambda_{\rm PSF}/p_{\rm in}$.\footnote{That is, all images simulated with the same $\lambda_*$ are processed using the same $\lambda_{\rm PSF}$, which results in significantly less computation time since IMCOM can apply a single solution to every sub-image in a dither set.  The reconstructed images will be unbiased, but some would have sub-optimal noise if our images contained noise.}

\begin{table}[p]
	\begin{center}
		\begin{tabular}{lccccc}
		\toprule
\emph{Simulation parameters} \\
			Source wavelength (\um) & $\lambda_*$ & 0.584 & 1.17 & 1.75 & 2.33\\
			Input pixel scale (\um) 		&$p_{\rm in}$& 10.0 & 10.0 & 10.0 & 10.0\\
			Input stamp width (pixels) 	&& 39 & 75 & 113 & 151\\
			Input stamp width (\um)	 	&& 390 & 750 & 1130 & 1510\\
			Number of random dithers 	&$N_d$& 30 & 8 & 5 & 2\\
\midrule
			Sampling Factor &$Q$  & 0.5 & 1.0 & 1.5 & 2.0\\
\midrule
\emph{Pipeline parameters} \\
			PSF model wavelength (\um) 	&$\lambda_{\rm PSF}$& 0.5 & 1.0 & 1.5 & 2.0\\
			Output pixel scale (\um) 	&$p_{\rm out}$& 2.5 & 5.0 & 7.5 & 10.0\\
			Output stamp width (pixels) 	&& 41 & 41 & 41 & 41\\
			Output stamp width (\um)		&& 102.5 &  205 & 307.5 & 410\\
			Input stamp border (\um)    	&& 10 & 20 & 45 & 60\\
\midrule
\emph{Reference parameters} \\			
			Reference pixel scale (\um) 		&$p_{\rm out}$& 2.5 & 5.0 & 7.5 & 10.0\\
			Reference stamp width (pixels) 	&& 151 & 151 & 151 & 151\\
			Reference stamp width (\um) 		&& 377.5 & 755 & 1132.5 & 1510\\
			Shear calibration &$P/2$ 		& 1.9 & 0.96 & 0.79 & 0.74\\
		\bottomrule
		\end{tabular}



        \caption{Parameters for our four sets of point source simulations. \emph{Simulation parameters} are used to render undersampled input images for IMCOM.  Each of these sets includes a range of ellipticities, with the sampling factor $Q$ set by the maximum ellipticity, $|e|=0.3$.  \emph{Pipeline parameters} are used to configure IMCOM and render the oversampled output images.  \emph{Reference parameters} are used to render oversampled reference images to which to compare the IMCOM output.  In the long wavelength case, where $Q=2.0$, the input images are already oversampled.  We can nevertheless combine images with IMCOM to test its performance in this regime.}
		\label{tab:simparams}
	\end{center}
\end{table}

\subsection{Baseline tests or ``pipeline error''} \label{sec:baseline}
In our first set of tests, we combine dithered images assuming that the relative offsets between the image coordinates are perfectly known.  
Since the images contain no simulated noise, the only differences between the IMCOM output images and the reference images will be due to IMCOM itself or the numerical accuracy of our simulated images and analysis. \reffig{f1} shows the differences in shear for the four cases summarized in \reftab{simparams}.  
The markers show the average bias over 100 simulations (random dither sets), and the solid lines show the standard deviations.  
The main takeaway is that in all cases, neither the bias nor the scatter is larger than $10^{-5}$, staying well below the $O(10^{-4})$ requirements for upcoming cosmic shear analyses.  
Similarly, \reffig{f2} shows that size errors are limited to $10^{-5}$, two orders of magnitude below the requirement for cosmic shear.  Consequently, we have shown that our settings for IMCOM and the rest of the pipeline are sufficient when applied to each of the simulation cases.  In particular, setting the tolerance on IMCOM's leakage objective to $U^{\rm max}_{\alpha} / C_{\alpha} = 10^{-7}$ is acceptable for weak lensing analysis (see \refsec{u7u8}).

At this stage, since these baseline errors are negligible, we do not find it useful to track down the cause of the observed trends in \reffig{f1} and \reffig{f2}.  
They are likely to be the result of our simulation or pipeline settings and not physically meaningful.
For instance, we may find a further reduction in the shear differences if we match our Gaussian weighting function to the centroids of the oversampled images to better than $10^{-4}$ pixels (see \refsec{moment}); however, that would be an unrealistic tolerance in practice since shot noise and other systematic effects would dominate the centroid measurement errors.  
In other words, the errors in in \reffig{f1} and \reffig{f2} are unlikely to be detectable in real data.  The different size errors when varying $e_1$ or $e_2$ is perhaps unsurprising since the calculations are sensitive to sub-pixel-scale effects, and $e_1$ is the ``plus'' polarization aligned with the pixel grid.

\Cfig{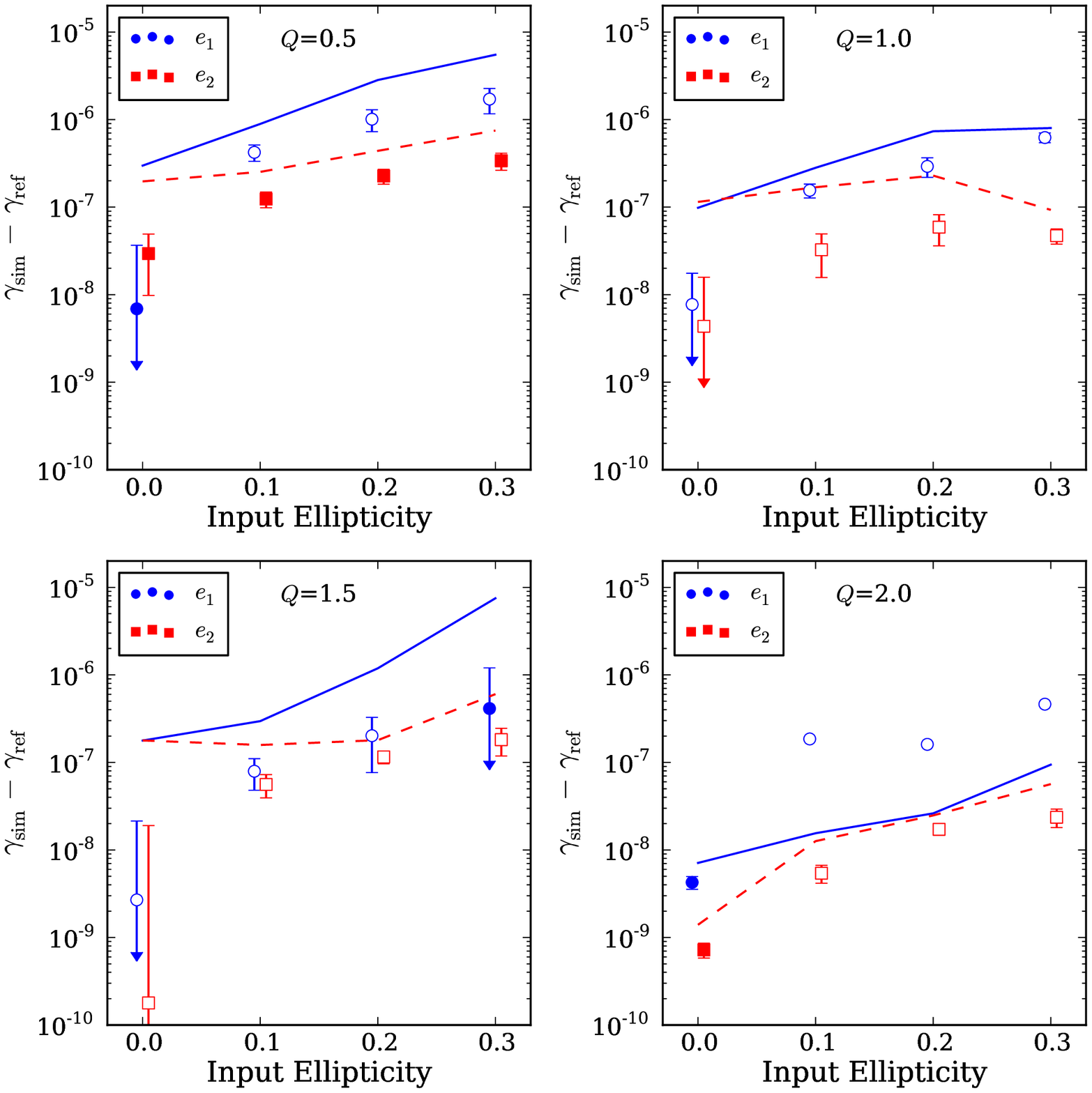}{Differences between shear measurements of oversampled reference images ($\gamma_{\rm ref}$) and images reconstructed by IMCOM ($\gamma_{\rm sim}$). 
For each sampling factor $Q$, the image model is sampled at the pixel scale $p$ to create input images for IMCOM and at the higher output rate to create reference images (see \reftab{simparams}).  Input ellipticity refers to the shape of the scaled Airy ($e_1$ or $e_2$ with the other fixed at zero), and the $\gamma$ plotted corresponds to the polarization of the markers (not an absolute value).
The markers (offset horizontally for clarity) show the average bias over 100 simulations, with closed and open symbols denoting positive and negative values, respectively.  Biases consistent with zero (within the $1\sigma$ error bar) have their lower error bars replaced by arrows.  Small error bars are occasionally hidden by the markers. The solid and dashed lines show the standard deviations for $\gamma_1$ and $\gamma_2$ measurements, respectively.  For upcoming large cosmic shear surveys, the total shear error budget is expected to be $\sim 2\times 10^{-4}$.
\emph{It is likely that the observed trends reflect the numerical accuracy of our simulated images and analysis, not the accuracy of IMCOM} (see \refsec{baseline}).}

\Cfig{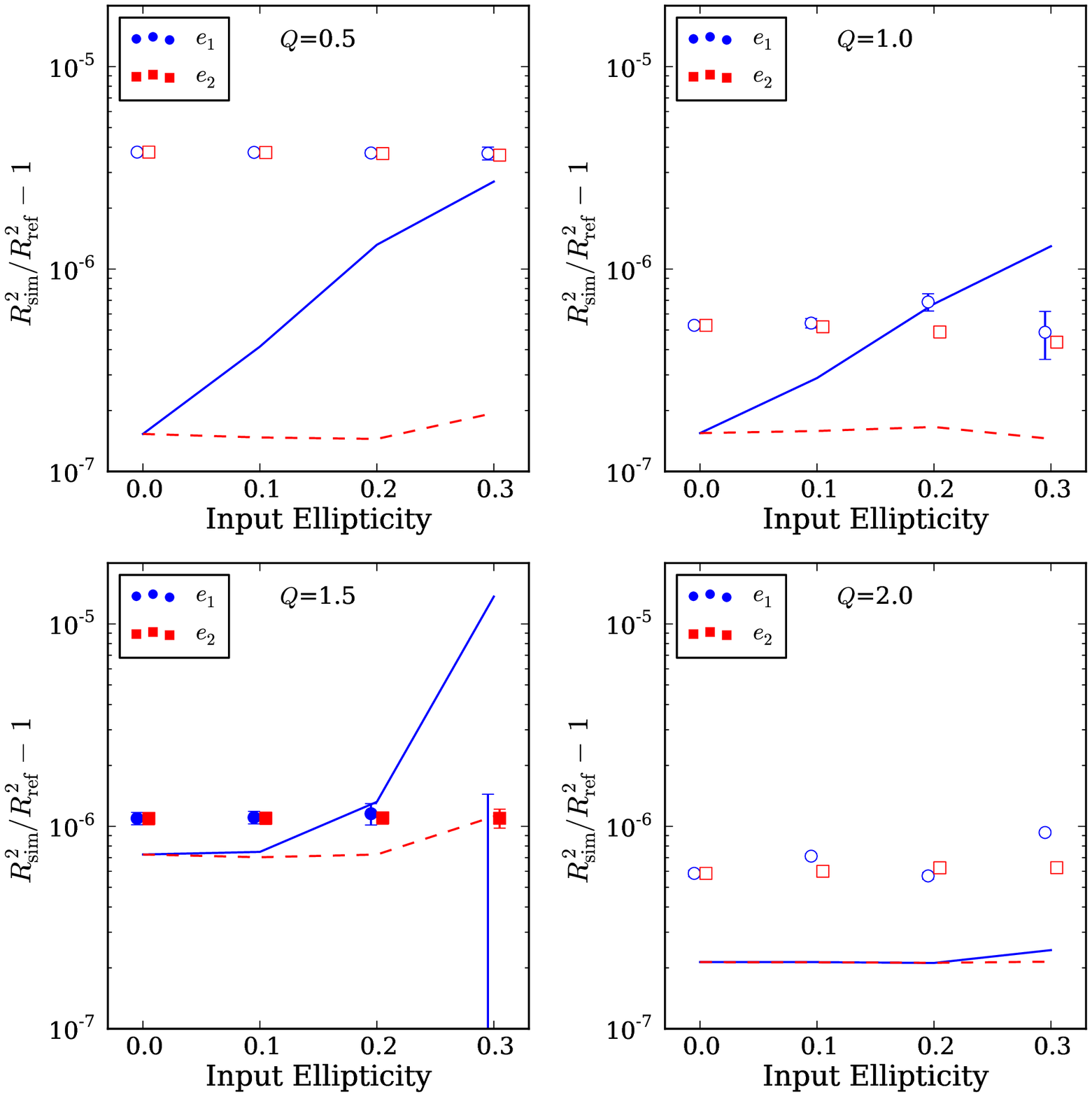}{Same as \reffig{f1} but comparing measurements of the fractional difference in the size, $R^2$. For upcoming large cosmic shear surveys, the total error budget for $dR^2/R^2$ is expected to be $\sim 10^{-3}$.
\emph{It is likely that the observed trends reflect the numerical accuracy of our simulated images and analysis, not the accuracy of IMCOM} (see \refsec{baseline}).}

\subsection{Image distortion due to errors in astrometric registration}
In our next set of tests, we look at image distortions due to errors in internal astrometric registration, i.e.~the measured dither positions of the input images.  In practice, the relative coordinates between input images would be determined by matching stellar positions.  There will be some uncertainty in the centroid measurement of each point source due to e.g. shot noise or detector systematics.  As described in \refsec{multistar}, we simulate such errors by displacing the centroids of the input images by random amounts without updating the positional information supplied to IMCOM.  The centroid displacements are drawn from a Gaussian distribution $(0,\sigma_d)$, measured in input pixels.  The $x$ and $y$ displacements are sampled independently. Thus, IMCOM knows the dither positions in either dimension with an uncertainty $\sigma_d$ relative to some fixed coordinate system.  In other words, the input images being combined are slightly misregistered with a scatter of $\sqrt{2}\sigma_d$ in the displacement between each pair of images.

The resulting shape measurement errors as a function of $\sigma_d$ are shown in \reffigs{f3}{f6}.  As one might expect, the bias and scatter in the shape measurement errors increase with $\sigma_d$.  As $\sigma_d\rightarrow 0$, we expect the errors to limit to the results in \reffig{f1} and \reffig{f2}, where no uncertainty was introduced.  In some cases, this limiting behavior is seen as a deviation from the measured power-law.  We reiterate that these deviations are likely due to numerical limitations in our simulation and pipeline, not physically meaningful effects.  We see no discernible differences between the various ellipticity cases, at least not in the limit of large $\sigma_d$, where the effect of astrometric registration error dominates.  The insensitivity of the shear errors to the image ellipticity implies that they are additive (as opposed to multiplicative) errors. Although we increased the number of simulations to 400 to reduce the error bars, many of the biases in the more undersampled cases are consistent with zero, indicating that their trends are dominated by statistical noise.
 
\Cfig{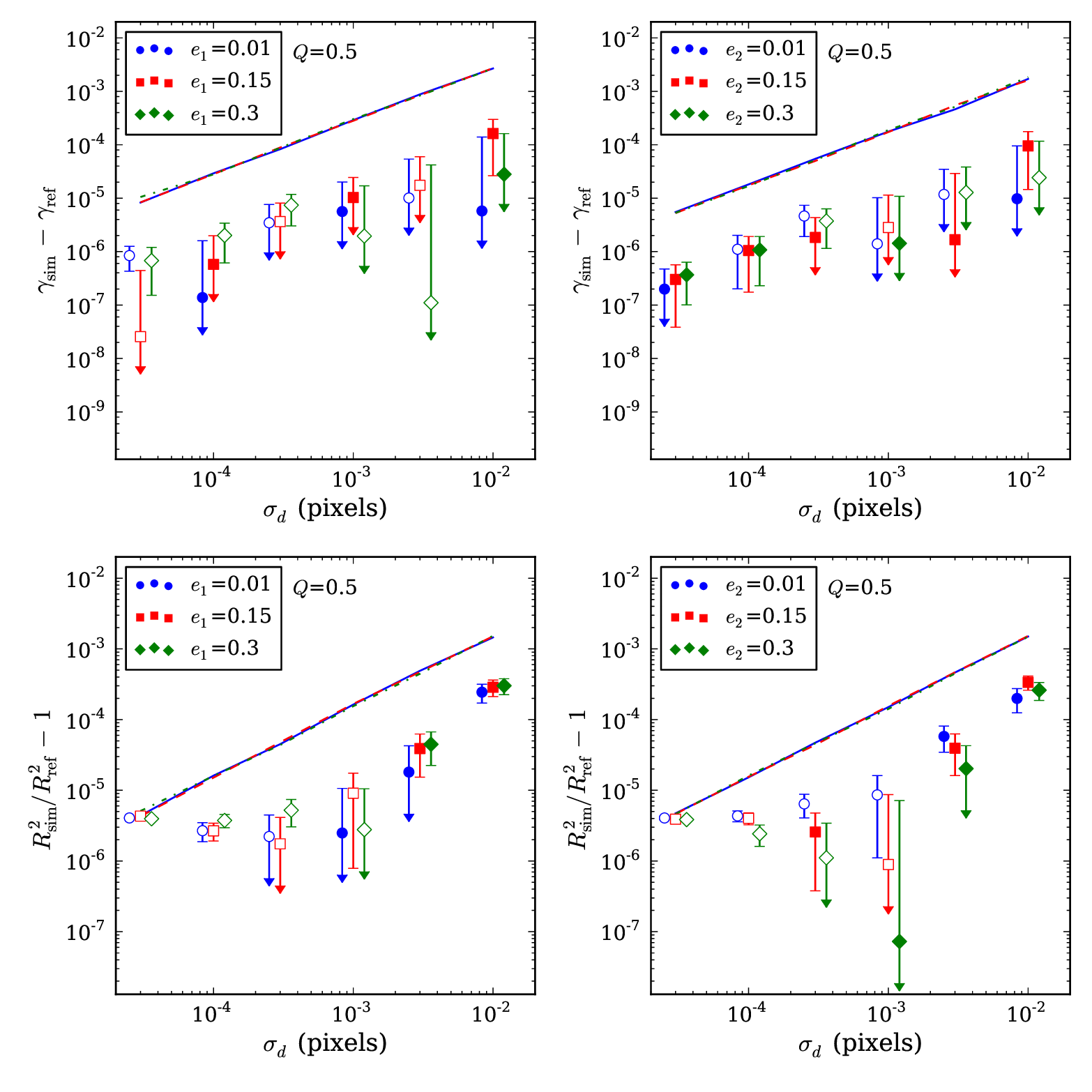}{TOP ROW: For our $Q=0.5$ case, the differences between shear measurements of oversampled reference images ($\gamma_{\rm ref}$) and images reconstructed by IMCOM ($\gamma_{\rm sim}$) with uncertain dither positions.  The relative coordinates between the input images (astrometric misregistration) have scatter in the $x$ and $y$ directions given by $\sigma_d$.  The left (right) panel shows various input ellipticities $e_1$ ($e_2$) with the other polarization fixed at zero.  This ellipticity refers to the shape of the scaled Airy function, and the $\gamma$ plotted corresponds to the polarization of the markers (not an absolute value).
The markers (offset horizontally for clarity) show the average bias over 400 simulations, with closed and open symbols denoting positive and negative values, respectively.  Biases consistent with zero (within the $1\sigma$ error bar) have their lower error bars replaced by arrows.  Error bars are occasionally small enough to be hidden by the markers. The solid, dashed, and dash-dotted lines show the standard deviations for $e_i=$ 0.01, 0.15, and 0.3, respectively.  BOTTOM ROW:  Same as the top row but comparing measurements of the fractional difference in the size, $R^2$.}

\Cfig{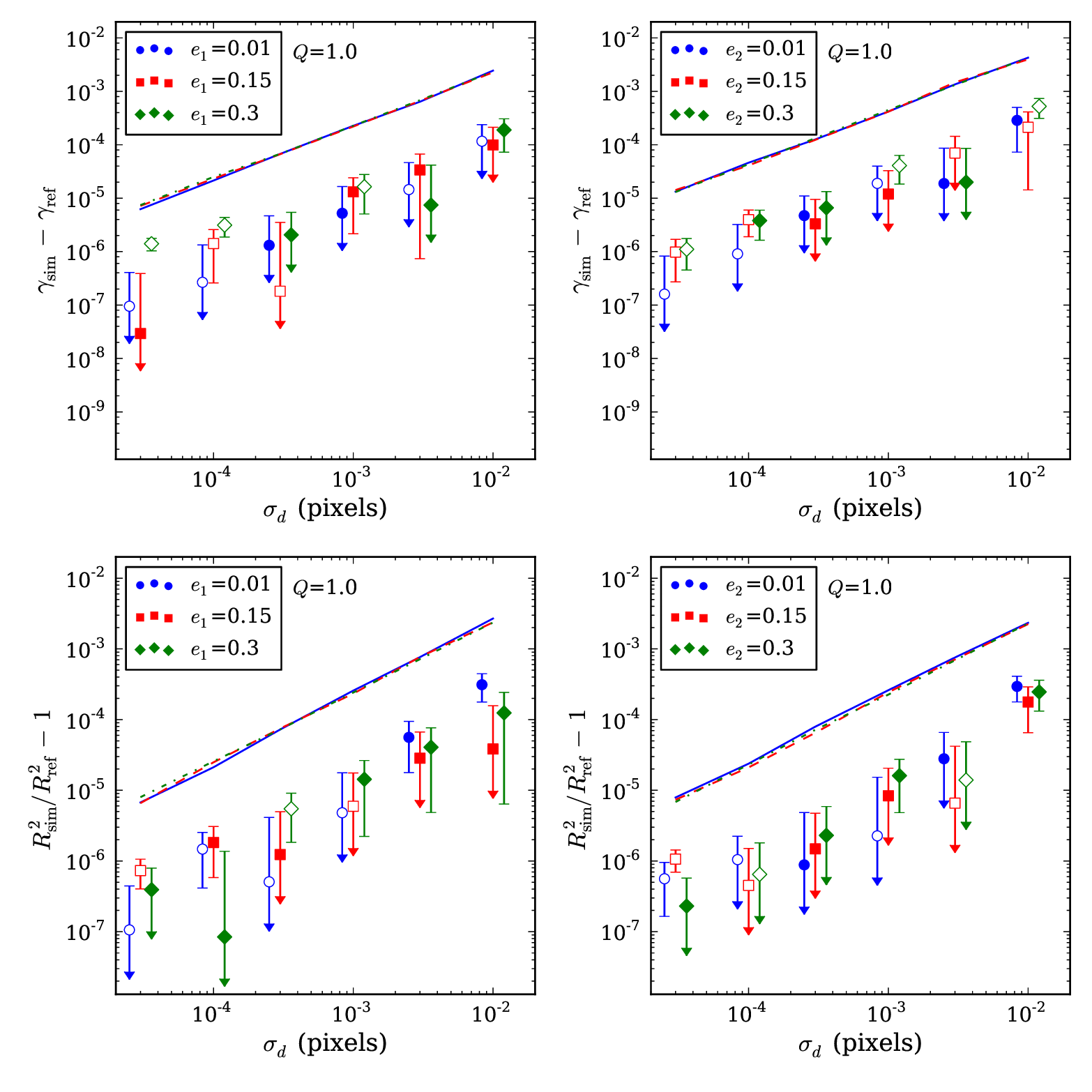}{Same as \reffig{f3} but for our $Q=1.0$ case.}
\Cfig{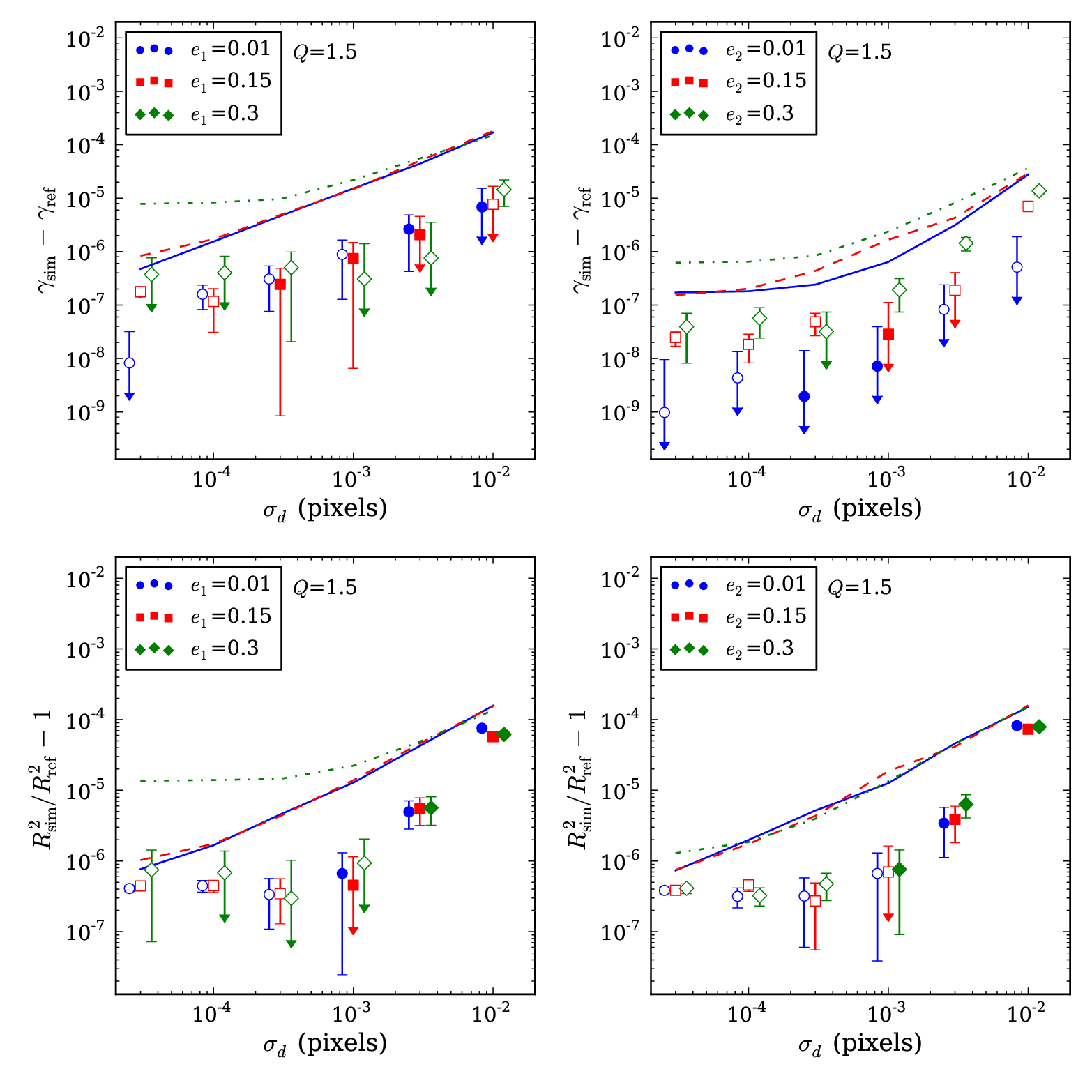}{Same as \reffig{f3} but for our $Q=1.5$ case.}
\Cfig{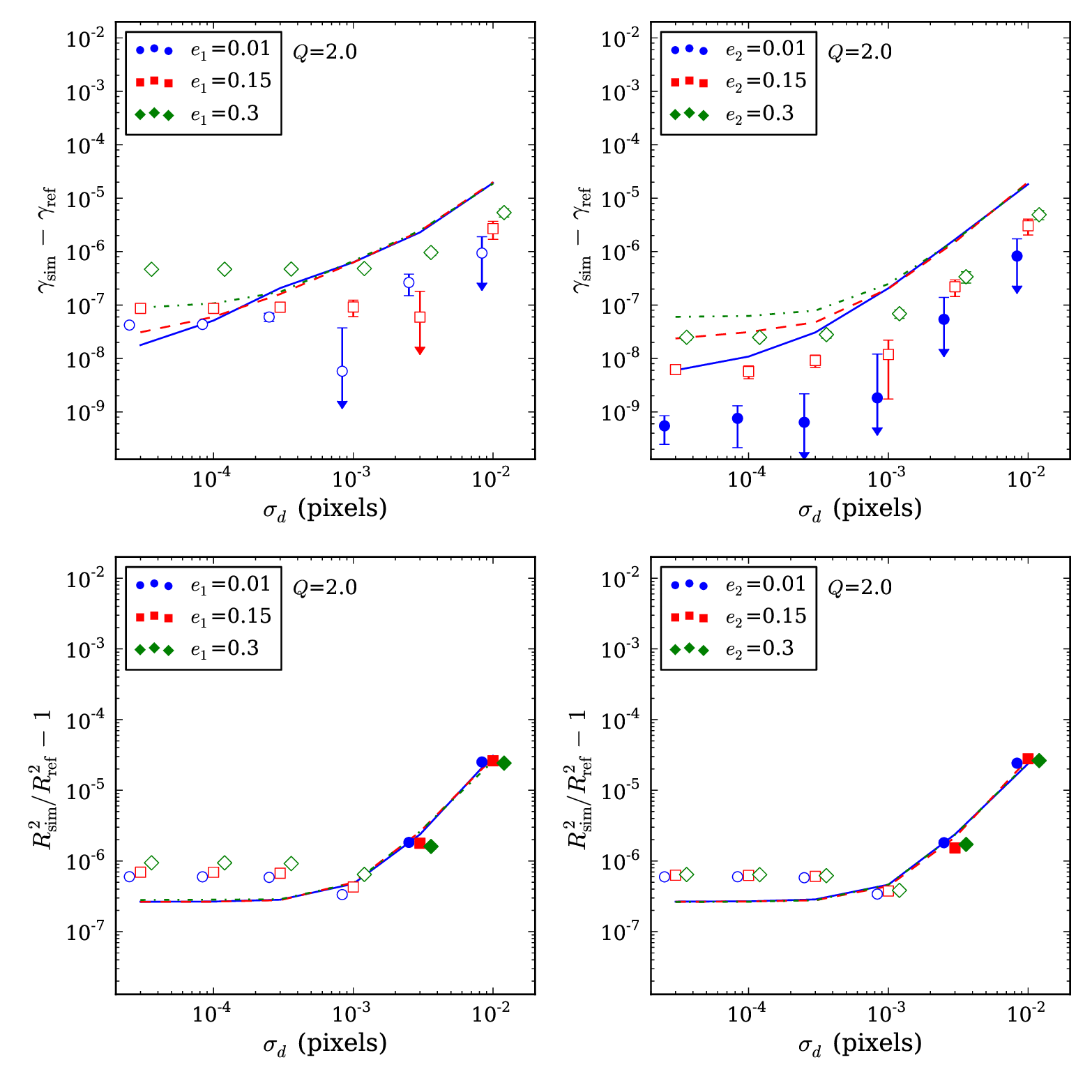}{Same as \reffig{f3} but for our $Q=2.0$ case.}

At large $\sigma_d$, we expect the effect of astrometric misregistration to be similar to the effect of astronomical seeing.  For instance, we expect the sign of the bias in $R^2$ to be positive.  The centroids of the misregistered input images are spread over a region of radius $\sim\sigma_d$, which can only make the combined image larger.  This expectation for the sign is confirmed by the data.  For the shear, the blurring of the image will tend to reduce the absolute value if the original image ellipticity was non-zero.  On the other hand, there is likely to be some asymmetry in the registration error of the small number of input images, which would induce a randomly oriented ellipticity.  The latter effect should be unbiased, therefore we expect to underestimate the shear.  Confirmation would require us to run more simulations in order to further reduce the error bars.

To put the size of $\sigma_d$ into context, we estimate the astrometric registration uncertainty of a WFIRST-AFTA pointing under pessimistic conditions.  
According to the Trilegal model v1.5
\citep{Girardi:2005ve}, the minimum number of single, bright stars that are far from
saturated in the 0.281
deg$^2$ WFIRST-AFTA pointing is estimated to be 620--720 at the South
Galactic Pole (depending on the filter). Here, ``bright stars'' means generating at least $10^4$ photo-electrons in the IR detector over a 184 s exposure, and ``far from saturated'' means generating no more than $5\times 10^4$ electrons in any one pixel.  For a given star, the statistical
error in the astrometry can be assessed from the Fisher matrix for the
flux and position of the star in a PSF model fit: this is 0.014 pixels RMS
per axis for the WFIRST-AFTA H-band PSF at the minimum source flux of $10^4$ electrons.
Thus, taking the minimum number of useable\footnote{We assume that stars falling near anomalous pixels are rare and are thrown out}
 stars in a pointing to be 620 and assuming the minimum signal-to-noise for each star, the  astrometric registration error for the full field should be no worse than $\sigma_d = 0.014*\sqrt{3/2}/\sqrt{620} = 6.9\times 10^{-4}$ pixels, where the factor of $\sqrt{3/2}$ accounts for the ``roll'' degree of freedom in the pointing.  Repeating the calculation using all non-saturated stars in the field and their individual signal-to-noise levels, we find $\sigma_d=7.6\times 10^{-5}$ pixels.  Much of this improvement comes from having several very bright stars in the field (signal-to-noise of 100-500).

Consider then \reffig{f4}, the scenario most similar to WFIRST-AFTA, which would combine 8 dithered images with $Q=1.08$ (although WFIRST-AFTA will include rotated dithers while we consider only translations).  Taking $\sigma_d=6.9\times 10^{-4}$ as a worst-case scenario, we see that all biases are well below the requirements for cosmic shear.  The standard deviations (solid lines) are perhaps more cause for concern, since e.g. they are $O(10^{-4})$ for shear.  Thus, any individual reconstructed image could have an appreciable shear bias, even though the average bias over all images is small.  Also, the biases of nearby sources will be correlated since they will be reconstructed based on the same dither measurement errors.  Those correlations could in turn bias measurements of the shear correlation functions and the cosmological parameters inferred from them.  The shape distortion effects of misregistration under these conditions may therefore be interesting for further study; however, we reiterate that this is a pessimistic scenario.

The $Q=1.5$ case in \reffig{f5} would be similar to WFIRST-AFTA in a longer wavelength filter (e.g. K-band).  Here, the shape distortions are an order of magnitude smaller than in the $Q=1.0$ case -- well out of range of weak lensing requirements, even for our pessimistic $\sigma_d$.  Although the longer wavelength would not be ideal for weak lensing due to the significantly larger PSF, we can be confident that image combination from 5 dithers will have negligible effect on shape measurement.  The $Q=2.0$ case in \reffig{f6} shows that there is also negligible distortion when combining 2 initially oversampled images.  Although this is not strictly necessary for accurate shape measurement, each additional co-added dither would reduce the noise covariance in the final image.  The $Q=0.5$ case in \reffig{f3} case demonstrates that unbiased shape measurements can be extracted from even extremely aliased images using a brute-force dithering approach.
This case is unrealistic for a weak lensing survey, particularly since many random dithers are needed to ensure that IMCOM routinely\footnote{The tolerance $U^{\rm max}_{\alpha} / C_{\alpha} = 10^{-7}$ can be met with fewer than 30 random dithers but with an increased failure rate.  See \refsec{multistar}.}
 reaches our specified tolerance on PSF distortion.  
 Such strongly undersampled PSFs are nevertheless useful in PPL detector characterization experiments, which use them to investigate sub-pixel detector effects or to achieve ultra-fine focus control.  We have shown here that image reconstruction will not be the limiting factor in the PPL pipeline so long as we have sufficiently accurate dither measurements. IMCOM is therefore a unique and effective tool for detector characterization.

\subsection{Extended Sources} \label{sec:extended}

In \refsec{setup}, we claimed that shape distortions for reconstructed point source images would constitute a worst-case scenario.  To support this claim, we repeated the $Q=1.0$ case (the case closest to a WFIRST-AFTA weak lensing survey) for extended sources.  Our extended source profile is the exponential function described in \refsec{sources}.  To approximate a typical galaxy in a weak lensing survey, we take the FWHM of the exponential to be equal to the FWHM of the Airy function in our PSF for $Q=1.0$.  Since the exponential FWHM is $1.3863\, r_0$, this corresponds to setting
\begin{equation}
r_0=\frac{1.0290}{1.3863}\lambda_* N_f \;.
\end{equation}
To create elliptical images, we shear the exponential profile before convolving with the PSF.  
Unlike in the point source case, there is no factor of $s$ increase to the spatial bandlimit of the elliptical input images.  Therefore, it is valid to set $\lambda_{\rm PSF}=\lambda_*$ for all images.  We run IMCOM on the extended sources using the same input PSF model from the point source case (with $\lambda_{\rm PSF}=1.0$\um).  Furthermore, the centroids of the extended sources are the same as those of the point sources: they use the same sample of random initial positions, random dithers, and random errors.  Therefore, IMCOM is applying the same transformation to the extended source images as to the point source images, and the results can be easily compared.  For shape measurement, we also use the same Gaussian weighting function as in the point source case ($\sigma=4.5$ output pixels).  We find that the shear calibration factor for the extended sources is $P/2 = 1.04$, as compared to 0.96 for the point source case (see \reftab{simparams}).

Comparing \reffig{f7} to \reffig{f4}, we find that shape distortions are generally smaller in the extended source case for the entire range of $\sigma_d$ tested.  This is true for both the bias and scatter and for both the shear and size measurements.  In particular, the scatter in the shear errors are at or below $10^{-4}$ even with the pessimistic astrometric registration error of $\sigma_d=6.9\times 10^{-4}$ pixels.  
Of course, shape measurement errors will vary with the size and profile of the extended sources; hence, it is difficult to estimate errors for a realistic survey without detailed knowledge of the morphologies of the galaxy source population.  
Regardless, our point source tests establish a useful upper limit on shape measurement errors from image combination.  

\Cfig{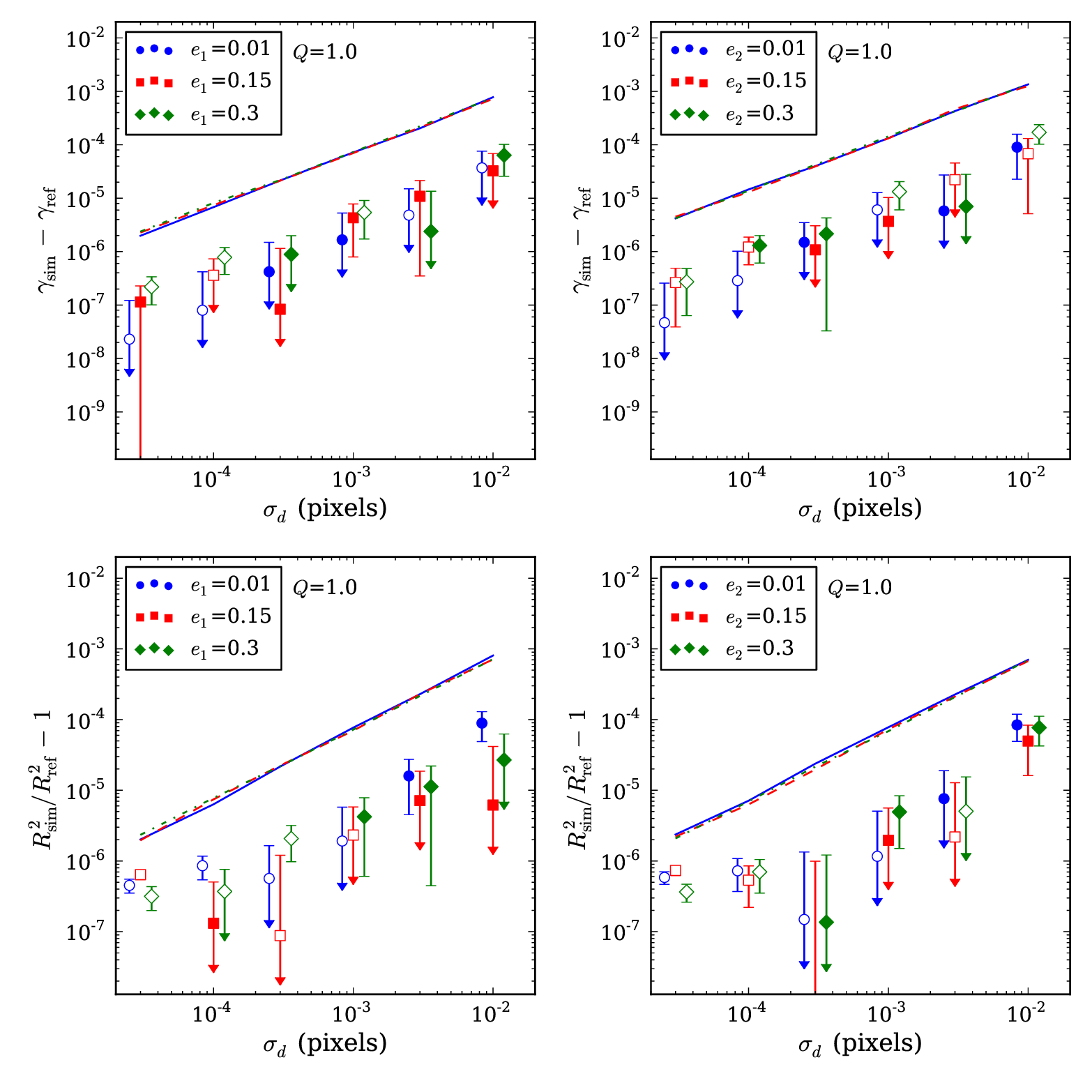}{Similar to \reffig{f4}, but instead of point sources, the underlying images are resolved exponential profiles convolved with the (non-sheared) PSF for the $Q=1.0$ case.  The ellipticities ($e_1,e_2$) in the legends refer to the shape of the scaled exponential function before PSF convolution.  The exponential FWHM is the same as the FWHM of the PSF for $|e|=0$.  The sample of random dither positions and errors are the same as those in the point source case in \reffig{f4}.  The shear and size measurement differences are generally smaller than in \reffig{f4} for all dither errors $\sigma_d$.}

\section{Conclusions} \label{sec:conclusions}

Large weak lensing surveys such as WFIRST-AFTA will need to accurately measure source shapes from aliased images.  The image distortions introduced by aliasing can be avoided by reconstructing oversampled images from multiple, dithered, raw images, but the reconstruction algorithm itself may also be a source of distortion.  We have investigated shape measurement errors due to image combination by comparing simulated images reconstructed with the IMCOM algorithm to simulated oversampled reference images.  The simulated images were point source images based on a simple PSF model which includes diffraction, charge diffusion, pixel response, and an artificial shearing (as from an elliptical pupil).  These images are generally more susceptible to pixel-scale effects than resolved images such as galaxies.

Over a range of ellipticities ($|e|\le 0.3$) and sampling factors ($0.5 \le Q \le 2.0$), we find that IMCOM creates negligible shear and size distortions (if any) when the relative offsets between the dithered input images are known, i.e. the input images can be precisely aligned.  
Our IMCOM settings included a tolerance on the ``leakage objective'' (error in the reconstructed PSF; see \refsec{u7u8}) of $U^{\rm max}_{\alpha} / C_{\alpha} = 10^{-7}$; we have therefore shown that running IMCOM with this tolerance is sufficient for cosmic shear analyses.
We also calculated shape measurement errors as a function of the astrometric uncertainty $\sigma_d$ in dither positions (or scatter in dithered image misalignment).  These results can be used to gauge the required astrometric precision of survey pointings given some tolerance for shape measurement error.  For a survey similar to the WFIRST-AFTA 2.4m telescope, in a pessimistic case assuming a low number of stars each with low signal to noise ($\sigma_d=5.6\times 10^{-4}$ pixels), we find that shape distortions due to dither misalignment are negligible on average but can have significant scatter ($\sim 10^{-4}$ for shear).  Much better pointing precision than this pessimistic case is likely to be achieved in practice; nevertheless, our results may be useful for determining weak lensing survey requirements.

IMCOM is an integral part of the analysis pipeline of the Precision Projector Laboratory (PPL), a joint project between NASA Jet Propulsion Laboratory and Caltech to study detector-induced distortions measured from images emulated in the lab.  Our results show that IMCOM is not a significant source of systematic bias in PPL experiments and that shape distortions in our reconstructed images can be controlled through precise measurements of the dither positions.  This supports the conclusions of an initial emulation study that was recently published \citep{Seshadri2013}.

\subsection*{Acknowledgements}
We thank E.~Jullo of Laboratoire d'Astrophysique de Marseille and V.~Velur for their contributions to the analysis pipeline.  This research was carried out at the Jet Propulsion Laboratory and California Institute of Technology, under a contract with the National Aeronautics and Space Administration. We are grateful to the following organizations and programs for their support of this effort: internal JPL Research and Technology Development (RTD) and DirectorÕs Research Development Fund (DRDF) programs; US Department of EnergyÕs (DOE) Supernova Acceleration Probe (SNAP) and Joint Dark Energy Mission (JDEM) projects; the NASA Wide Field IR Survey Telescope (WFIRST) and Joint Dark Energy Mission (JDEM) project offices.  CS acknowledges support from a NASA Postdoctoral Program Fellowship from Oak Ridge Associated Universities.  BR acknowledges support from the European Research Council in the form of a Starting Grant with number 240672.  JPL is run by the California Institute of Technology under a contract for NASA.  Thanks also to our anonymous referee for improvements to this manuscript.



\let\clearpage\relax  

\appendix
\twocolumn
\section{Appropriate Models for the PSF inputs to IMCOM}\label{sec:ggamma}
    
Processing the test images with IMCOM requires models for the system
PSF $G_i$ and target PSF $\Gamma$ (see \refsec{imcom}).  In
order for the IMCOM algorithm to find a genuinely optimal solution to
the image combination problem $G_i$ should be as close to the true
input PSF as possible.  It is worth briefly commenting more on how
well-motivated choices are made for $G_i$ and $\Gamma$, given that
both must inevitably be made without perfect information about the true
PSF for any system in practice.

If the model for $G_i$ \emph{underestimates} the bandlimit in the
images $I_i$ (equation \ref{eq:I}), or in other ways spuriously sets
to zero spatial frequencies which are in fact contained in $I_i$, this
will have a serious negative impact on the accuracy of the output
$H_{\alpha}$ as a representation of the expression in \eqref{eq:Ja}.
$H_{\alpha}$ will in general then contain (possibly aliased) spatial
frequency modes which were not correctly treated in the solution for
$T_{\alpha i}$.  These will cause distortions and defects in
$H_{\alpha}$ that are not described by either $U_{\alpha}$ or $\Sigma_{\alpha
  \alpha}$.  Underestimating the bandlimit in the model for $G_i$
therefore represents a primary mode of failure for the entire IMCOM
approach.  Happily, the bandlimit for any telescope image is usually
\emph{extremely} well determined by the diameter of the primary mirror
and the wavelength at the blue edge of the observing filter.

If instead the model for $G_i$ \emph{overestimates} the bandlimit in
the image $I_i$ the effects are more benign.  The solution $T_{\alpha
  i}$ will not in general be optimal in terms of its noise properties,
and more dithers might be required for control over $U_{\alpha}$ than
for a perfect model of $G_i$.  However, the output image $H_{\alpha}$
will not be distorted in any way not already characterized by
$\Gamma$, the output leakage objective $U_{\alpha}$, and noise
covariance matrix $\Sigma_{\alpha \beta}$.  Estimates of the system
PSF from images of stars in the output image, for example, will be
unbiased and free from hidden distortions.

Other errors that may affect $G_i$, such as uncertainties on the
precise shape of the PSF due to interpolation from a finite number of
stars, will also impact the degree to which the solution $T_{\alpha
  i}$ is optimal.  However, this can be very effectively mitigated in
practice by choosing $\Gamma = G_i$.  Under this choice IMCOM will
successfully limit unwanted changes in the PSF
to the degree specified by $U_{\alpha}$, even if $G_i$ is
only an approximate model of the PSF, provided of course that $G_i$ does not
wholly filter out modes that are in fact present in the images $I_i$
(see above).

Therefore, in the tests presented we always i) choose
$\Gamma=G_i$; and ii) choose for the model $G_i$ to be bandlimited at
or above the true bandlimit present in the input signal.  Furthermore,
for these tests of a multipurpose image recombination pipeline we will
not include the effects of pixellization in $G_i$ (the function
$I_{\rm pixel}$ in \refsec{sources}).  As shown in equation
\eqref{eq:ftpixel} the theoretical model for this function has zero
crossings at $u_1, u_2 = N/p$ in the Fourier domain, where $N$ is an
integer.  One purpose of an image processing pipeline that uses IMCOM
on undersampled images is in the field of detector characterization,
and here the theoretical model of equation \eqref{eq:ftpixel} may only
be approximately correct.  By including $I_{\rm pixel}$ in the model
for $G_i$ these spatial modes would be forcefully assumed to be zero
when they may in fact be present.  The visible effects of pixellization will
still be represented in the output image $H_{\alpha}$ as they are
present in the input, but assumptions about these effects will not be
made to optimize the image reconstruction.

Section \ref{sec:sources} describes the PSF due to a simple model of
optical diffraction, charge diffusion and pixellization.  For our
tests we model $G_i$ and $\Gamma$ as the convolution of $I_{\rm
  optical}(r; \lambda_{\rm PSF} N_f)$ and $I_{\rm cd}(r; \sigma_{\rm cd})$, for
$N_f=10$ and $\lambda_{\rm PSF}$ as given by Table \ref{tab:simparams}.  We
purposefully omit $I_{\rm pixel}$ from $G_i$ or $\Gamma$, although the
pixel integration will be correctly applied to the input images via
the full expression \eqref{eq:ftsource}.

As simulated PSF ellipticities are introduced by applying a linear
coordinate shear to the pupil (see \refsec{sources}) this in
fact increases the bandlimit in the minor axis direction and reduces
it along the major axis direction.  To ensure that the sheared
simulated stars remain fully bandlimited for the chosen $G_i$ and
$\Gamma$ we in fact adopt a somewhat redder wavelength, $\lambda_*=s\lambda_{\rm PSF}$,
specifically for generating the input images $I_i$.  Here, $s$ is given by the scaling factor in \refeq{scalingfactor}.  As discussed, since $G_i$ now
underestimates the bandlimit of the input images, the image
reconstruction $H_{\alpha}$ will not be strictly optimal, but it should be unbiased in principle.  

\bibliography{shapes} 
\bibliographystyle{apj}

\end{document}